\documentclass[11pt,a4paper]{article}
\pdfoutput=1
\usepackage{jcappub}
\usepackage{multirow}
\usepackage{amssymb}

\newcommand{\fett}[1]{\boldsymbol{#1}}

\newcommand{\dd}{{\rm{d}}}
\newcommand{\ii}{{\rm{i}}}

\newcommand{\be}{\begin{equation}}
\newcommand{\ee}{\end{equation}}

\newcommand{\Nmb}{{{\rm Nm}\to{\rm Nb}}}

\definecolor{darkred}{rgb}{0.5,0,0}
\definecolor{darkgreen}{rgb}{0,0.6,0}
\definecolor{darkblue}{rgb}{0,0,0.5}

\newcommand{\vek}[1] {{\bf #1} }
\renewcommand{\vec}{\vek}

\newcommand{\CLASS}{{\sc class}}

\newcommand{\inspire}[1]{\href{http://inspirehep.net/search?p=find+J+#1}
 {{\color{black}[{\color{blue} {\small in}SPIRE}]}}}
\newcommand{\book}[1]{\href{http://inspirehep.net/search?p=#1}
 {{\color{black}[{\color{blue} {\small in}SPIRE}]}}}

\newcommand{\inspired}[1]{\href{http://inspirehep.net/search?p=#1}
 {{\color{black}[{\color{blue} {\small in}SPIRE}]}}}

\newcommand{\deltaNB}{\delta^\text{Nb}}
\newcommand{\velNB}{\fett{v}^\text{Nb}}
\newcommand{\gammaNB}{\gamma^\text{Nb}}

\begin{document}

\title{Relativistic Interpretation of Newtonian Simulations for \\ Cosmic Structure Formation}

\date{\today}

\author[a]{Christian Fidler,}
\emailAdd{christian.fidler@uclouvain.be}

\author[b]{Thomas Tram,}
\emailAdd{thomas.tram@port.ac.uk}

\author[c,b]{Cornelius Rampf,}
\emailAdd{rampf@thphys.uni-heidelberg.de}

\author[b]{Robert Crittenden,}
\emailAdd{robert.crittenden@port.ac.uk}

\author[b]{Kazuya Koyama}
\emailAdd{kazuya.koyama@port.ac.uk}

\author[b]{and David Wands}
\emailAdd{david.wands@port.ac.uk}

\affiliation[a]{Catholic University of Louvain - Center for Cosmology, Particle Physics and Phenomenology (CP3) 2, Chemin du Cyclotron, B--1348 Louvain-la-Neuve, Belgium}
\affiliation[b]{Institute of Cosmology and Gravitation, University of Portsmouth, Portsmouth PO1 3FX, United Kingdom}
\affiliation[c]{Institut f\"ur Theoretische Physik, Universit\"at Heidelberg, Philosophenweg 16, D--69120 Heidelberg, Germany}

\abstract{
The standard numerical tools for studying non-linear collapse of matter are Newtonian $N$-body simulations. Previous work has shown that these simulations are in accordance with General Relativity (GR) up to first order in perturbation theory, provided that the effects from radiation can be neglected. In this paper we show that the present day matter density receives more than 1$\%$ corrections from radiation on large scales if Newtonian simulations are initialised before $z=50$. We provide a relativistic framework in which \emph{unmodified} Newtonian simulations are compatible with linear GR even in the presence of radiation. Our idea is to use GR perturbation theory to keep track of the evolution of relativistic species and the relativistic space-time consistent with the Newtonian trajectories computed in $N$-body simulations. If metric potentials are sufficiently small, they can be computed using a first-order Einstein--Boltzmann code such as \CLASS{}. We make this idea rigorous by defining a class of GR gauges, the \emph{Newtonian motion} gauges, which are defined such that matter particles follow Newtonian trajectories. We construct a simple example of a relativistic space-time within which unmodified Newtonian simulations can be interpreted.
}

\maketitle   

\flushbottom
\section{Introduction}
\label{Introduction}

In recent years, the $\Lambda$CDM cosmological model has emerged as a successful framework that can reproduce a wide range of cosmological observations (e.g., \cite{Tegmark:2003ud,Suzuki:2011hu,Ade:2013zuv}). According to this model, we live in a Universe with an energy density
that is currently dominated by a cosmological constant ($\Lambda$) and a cold dark matter (CDM) component. 
Today, baryons and radiation provide relatively small contributions to the overall energy budget, yet these components played
a major role in the early dynamical evolution of our Universe.

Across a wide range of scales, cosmological structure formation is principally the result of gravitational instability from primordial seed perturbations.
The evolution of the Universe is thus governed by a coupled set of Einstein--Boltzmann equations.
Gravitational collapse is highly non-linear, and investigating the full Einstein--Boltzmann system in full non-linearity remains an open problem.
Instead, most simulations are based on the simpler Newtonian equations of motion, which assumes that Newtonian dynamics is the correct small scale limit of General Relativity (GR),
but important relativistic corrections to Newtonian results are expected on large scales \cite{Bruni:2013qta,Rigopoulos:2014rqa,Bertacca:2014dra,Villa:2015ppa,Christopherson:2015ank,Valkenburg:2015dsa}.

The standard numerical tools for studying non-linear collapse are Newtonian $N$-body simulations (see, e.g., \cite{Springel:2005mi,Teyssier:2001cp,Hahn:2015sia}).
In a previous paper~\cite{Fidler:2015npa} 
we have shown that a consistent relativistic interpretation of these simulations, up to first order in perturbation theory, can be constructed when the impact of radiation can be safely neglected. 
We performed a fully relativistic analysis of the multi-component system in the $N$-body gauge, and obtained the following relativistic fluid equations for the CDM component valid to first order 
\begin{subequations} \label{eqs:introN-body-gauge}
\begin{align}
\nabla^2 \Phi &=  - 4\pi G  a^2 \smash{\sum_\alpha} \bar \rho_\alpha \deltaNB_\alpha  \,, & \label{Poisson:Nbody} \\
\partial_\tau \delta^{\rm Nb}_{\rm cdm} + \nabla\! \cdot \! \velNB_{\rm cdm} &=  0 \,, &\text{($N$-body gauge)} \\
\left[ \partial_\tau + {\cal H} \right]  \velNB_{\rm cdm} \hspace{0.015cm}&= \nabla \Phi +  \nabla  \gammaNB \,,   & \label{Euler:Nbody}
\end{align}
\end{subequations}
where the superscript Nb refer to $N$-body gauge quantities. 
These equations are the relativistic versions of the Poisson, continuity and Euler equation respectively. 
They are formally identical to the linearised Newtonian fluid equations for CDM, apart from a relativistic correction term, $\nabla  \gammaNB$ in the Euler eq.\,(\ref{Poisson:Nbody}) (defined below in eq.\,(\ref{def:gammaFourier})), and the fact that in the multi-fluid case the Poisson equation (\ref{Poisson:Nbody}) is sourced by all the fluid species (labelled by the index $\alpha$), not just CDM.
As shown in ref.~\cite{Fidler:2015npa} the relativistic correction, $\nabla  \gammaNB$, becomes negligible at late times in $\Lambda$CDM and so does the impact of the relativistic components in the Poisson equation. 

The above analysis implies that Newtonian simulations are in one-to-one correspondence with GR at first order, provided that one uses initial conditions given in the $N$-body gauge and initialises the Newtonian simulation when the effect of radiation is negligible, i.e., at sufficiently late times. However, one would also like to begin the Newtonian simulation in a regime where linear perturbation theory is still valid on all scales of interest, i.e., at early times.  

One possible solution is to modify Newtonian  
codes to include the effect of GR and radiation~\cite{Adamek:2013wja,Adamek:2015eda,Hahn:2016roq}. 
However a great deal of time and effort has gone into optimising 
codes for the Newtonian system and they are not easily adapted to include relativistic components.
Instead we propose a different strategy that does not involve any modifications of existing Newtonian simulations and identify a tailor-made gauge (more precisely, a class of gauges) in GR that by definition absorbs the relativistic corrections in equation~\eqref{eqs:introN-body-gauge} within the coordinates of that gauge.  We call this a \emph{Newtonian motion} gauge.  
Specifically, the first-order CDM fluid equations in such a Newtonian motion (superscript Nm) gauge can be expressed as
\begin{subequations} \label{eqs:introNM-gauge} 
\begin{align}
  \nabla^2 \Phi^{\rm N} &=  - 4\pi G  a^2  \bar \rho_{\rm cdm} \delta_{\rm cdm}^{\rm N}   \,, &  \\ 
  \partial_\tau \delta_{\rm cdm}^{\rm N} + \nabla\! \cdot \!\fett{v}^\text{Nm}_{\rm cdm} &=  0 \,, &\text{(Newtonian motion gauge)}  \\
 \left[ \partial_\tau + {\cal H} \right] \fett{v}^\text{Nm}_{\rm cdm} \hspace{0.015cm}&= \nabla\Phi^{\rm N} \,, &
 \end{align}
 \end{subequations}
where the superscript N refers to quantities which are solved for in a Newtonian simulation. This set of equations is thus identical (up to first order) to the set of equations that Newtonian simulations solve. However, there is a crucial yet hidden difference between the GR equations~\eqref{eqs:introNM-gauge} and the ones solved in Newtonian simulations: the space-time in the Newtonian motion gauge is dynamically evolving precisely to accommodate the use of Newtonian equations of motion.

The dynamical change in the GR space-time is reflected in the full Newtonian motion gauge metric potentials. We are then able to add GR corrections on top of the unmodified Newtonian simulations as a kind of post-processing. In this paper we explicitly discuss how Newtonian simulations could be used in combination with a Newtonian motion gauge to give a relativistic interpretation.

This paper is organised as follows. 
In section \ref{sec:notation} we introduce the notation used for matter and metric perturbations. 
We introduce the relativistic equations in a general gauge in section \ref{radnbody}, review the $N$-body gauge and 
present the class of Newtonian motion gauges.  
In section~\ref{NMgauge} we discuss the residual gauge freedom and define a specific Newtonian motion gauge that can be viewed as an extension of the $N$-body gauge in the presence of radiation, while we discuss other gauge choices in appendices~\ref{app:A} --  \ref{app:tom}. In section~\ref{sec:appN-bodySim} we explicitly show how our method can be used in combination with a Newtonian simulation to derive the full relativistic results in either the Newtonian motion or the $N$-body gauge.  
We conclude in section~\ref{sec:conclusions}.

\section{Notation and conventions}\label{sec:notation}
Here we briefly review the notation and conventions we adopt.  Those familiar with the general framework of cosmological perturbation theory \cite{Bardeen:1980kt, Kodama:1985bj, Hu:2004xd, Malik:2008im} may wish to skip directly to the next section. 

We define metric perturbations on a homogeneous, isotropic and flat background space, with the cosmic scale factor $a$, in an arbitrary gauge:
\begin{align}
\label{eq:metric}
  \begin{split}
	g_{00} &= -a^2 (1+2A) \,, \\ 
	g_{0i} &= -a^2 B_i \,, \\
	g_{ij} &= a^2 \left[ \delta_{ij} \left( 1 +2 H_{\rm L} \right) - 2 H_{{\rm T}\,{ij}} \right] \,,
  \end{split}
\end{align}
with the time perturbation $A$, the shift $B_{i}$, and the trace and trace-free scalar perturbations of the spatial metric which are denoted by $H_{\rm L}$ and $H_{{\rm T}\,ij}$ respectively. 
We use the conformal time $\tau$ defined by $a \dd \tau \equiv \dd t$ where $t$ is the proper time in the unperturbed background. Partial derivatives with respect to conformal time are denoted with $\partial_\tau$ or an overdot. 

The background expansion is governed by the usual Friedmann equations.
The matter and radiation content is characterised by the stress-energy tensor 
\begin{align} 
 \begin{split}
	T^{0}_{\phantom{0}0} &= - \sum_\alpha \bar \rho_\alpha \left( 1  + \delta_\alpha \right) \equiv - \bar \rho \left( 1  + \delta \right) \,, \\
	T_{{\phantom{0}}0}^i &= - \sum_\alpha (\bar \rho_\alpha +\bar p_\alpha ) v_\alpha^i \equiv - (\bar \rho +\bar p ) v^i \,, \\ 
	T^{i}_{\phantom{i}j} &= \sum_\alpha (\bar p_\alpha+\delta p_\alpha ) \delta^i_j + \bar p_\alpha \Pi^{i}_{\phantom{i}j_\alpha} \equiv (\bar p+\delta p ) \delta^i_j + \bar p \Pi^{i}_{\phantom{i}j}  \,,
 \end{split} \label{Tmunu}
\end{align}
where the summation $\sum_\alpha$ runs over all species present in the Universe, i.e.,
 CDM, baryons, photons and neutrinos. 
Quantities without subscript will refer to totals as defined above.
Furthermore, $\bar \rho$ denotes the homogenous background density, $\delta= (\rho - \bar \rho) / \bar \rho$ the density contrast,
$\overline p$ the background pressure and its perturbation $\delta p$, and $\Pi^{i}_{\phantom{i}j}$ is the anisotropic stress of radiation.

We will often work in Fourier space, using 
an eigenmode expansion of the Laplacian operator
\be
  \nabla^2 S = - k^2 S \,.
\ee
For a spatially flat background Universe (as employed in the present paper), one choice of eigenmodes are the plane waves $S = \exp(\ii \fett{k} \cdot \fett{x})$. 
Considering only scalar perturbations, any vector $V_i$ (e.g., the velocity) and any trace-free tensor $T_{ij}$ (e.g., the perturbation $H_{{\rm T}\,ij}$) may be expanded using
\begin{align}
  V_i \equiv  - \ii \hat k_i V \,, \qquad \quad T_{ij} \equiv \left(  \delta_{ij}/3 - \hat k_i \hat k_j \right) T\,,
\end{align}
where $\hat k_i \equiv k_i/k$, with $k \equiv |\fett{k}|$.

From the general metric~(\ref{eq:metric}), we can define a specific gauge by choosing spatial and temporal coordinate transformations
\be
\label{def:generalgaugetrafo}
  \fett{x} = \fett{\tilde{x}} + \fett{L} \,, \qquad \tau = \tilde{\tau} + T \,.
\ee 
where for a scalar spatial coordinate transformation $\fett{L}=-k^{-1} \nabla L$.
Under this transformation, the metric potentials change at first order as:
\begin{subequations}
\begin{align}
 \label{eq:generalgaugetrafoA}
	A &= \tilde{A} -\dot{T} - \mathcal{H}T\,, \\ 
\label{eq:generalgaugetrafoB}
	B &= \tilde{B} + \dot{L} + k T\,, \\ 
\label{eq:generalgaugetrafoHL}
	H_{\rm L} &= \tilde{H}_{\rm L} - \frac k 3 L - \mathcal{H}T\,, \\ 
\label{eq:generalgaugetrafoHT}
	H_{\rm T} &= \tilde{H}_{\rm T} + k L\,,
\end{align}
\end{subequations}
while the matter fields change to:
\begin{subequations}
\begin{align}
 \label{eq:generalgaugetraforho}
	\bar{\rho}\delta &= \bar{\rho} \tilde{\delta} - \dot{\bar{\rho}} T\,, \\ 
\label{eq:generalgaugetrafop}
	\delta p &= \tilde{\delta p} - \dot{\bar{p}} T\,, \\ 
\label{eq:generalgaugetrafov}
	v &= \tilde{v} + \dot{L}\,, \\ 
\label{eq:generalgaugetrafoPi}
	\Pi &= \tilde{\Pi}\,.
\end{align}
\end{subequations}

It is useful in the Einstein equations to work with the gauge-invariant metric potentials (corresponding to the scalar metric potentials in the Poisson gauge) \cite{Bardeen:1980kt,Hu:2004xd}
\begin{subequations}
\begin{align}
\label{def:Psi}
\Psi &\equiv A+ {\cal H} k^{-1} \left( B- k^{-1}\dot H_{\rm T} \right) +k^{-1} \left( \dot{B}- k^{-1}\ddot H_{\rm T} \right) \,,
\\
\label{def:Bardeen}
 \Phi &\equiv H_{\rm{L}} +  \frac 1 3 H_{\rm{T}} + {\cal H} k^{-1} \left( B- k^{-1}\dot H_{\rm T} \right) \,.
\end{align}
\end{subequations}
We will also frequently make use of the gauge-invariant comoving curvature perturbation, defined as \cite{Hu:2004xd} 
\be \label{def:curvature}
 \zeta \equiv H_{\rm L} + \frac 1 3 H_{\rm T} + \mathcal{H} k^{-1}(B-v) \,.
\ee


\section{Relativistic interpretation of Newtonian simulations}\label{radnbody}

In this section we first review the linearised Einstein--Boltzmann equations governing the evolution of our multi-component Universe. We then briefly review the $N$-body gauge \cite{Fidler:2015npa} to quantify the impact of radiation on the matter evolution in this gauge. Finally we propose a novel gauge choice, that makes it possible to give a consistent relativistic interpretation of Newtonian dark matter trajectories calculated in Newtonian simulations even in the presence of radiation.

\subsection{Einstein and fluid equations in a general gauge}

The Einstein equations at first perturbative order in an arbitrary gauge are \cite{Hu:2004xd}
\begin{subequations}\label{eq:EEfull}
 \begin{align}
 	4 \pi G a^2 \left[ \bar \rho \delta + 3 {\cal H}  \left( \bar \rho+ \bar p \right) k^{-1} \left( v - B \right) \right] &= k^2 \Phi  \,, \label{eq:EE2} \\ 
 	k^2 \left( A + H_{\rm{L}} + \frac 1 3 H_{\rm T} \right) - \left[ \partial_\tau + 2{\cal H} \right] \left( \dot{H}_{\rm{T}}-kB \right)  &= -8 \pi G a^2 \bar p \Pi \,, 
	\label{eq:EEA} \\
 	4 \pi G a^2 \left( \bar \rho + \bar p \right) k^{-1} \left( v -B \right) &= {\cal H} A -\dot{H}_{\rm{L}} - \frac 1 3 \dot{H}_{\rm T} \,, \label{eq:EEB} \\
	\left( \partial_\tau +4{\cal H} \right) \left( \bar\rho+ \bar p \right) k^{-1} \left(v-B \right) &= \delta p - \frac{2}{3} \bar p \Pi + \left(\bar \rho + \bar p\right) A \,,\label{eq:Elast}
 \end{align} 
where the gauge-invariant Bardeen potential is defined in eq.~(\ref{def:Bardeen}).
These equations together with the Boltzmann equations for the various species in a general gauge form a closed set of equations.
The CDM component ($p_{\rm cdm} \equiv 0$ and $\Pi_{\rm cdm}\equiv0$) is completely described by its density and velocity, so the Boltzmann equation reduces to the relativistic continuity and Euler equations, which are respectively 
\begin{align} 
\dot{\delta}_{\rm{cdm}} + k v_{\rm cdm} &= - 3 \dot H_{\rm L} \,, \label{eq:NMconti} \\  
\left[ \partial_\tau + {\cal H} \right] \left(v_{\rm{cdm}}-B \right) &= k A \,. 
\end{align}
\end{subequations}
Using the Einstein equation~(\ref{eq:EEA}) we can simplify the dark matter Euler equation 
\be \label{eq:NMvelo}
 \left[ \partial_\tau + {\cal H} \right] v_{\rm cdm} = -k ( \Phi +  \gamma ) \,,
\ee
with the Bardeen potential $\Phi$ and a relativistic correction $\gamma$ describing the forces acting on the dark matter particles, given by 
\be \label{def:gammaFourier}
 - k^2  \gamma \equiv  \left( \partial_\tau + {\cal H} \right) \dot{ H}_{\rm T} - 8\pi G a^2 \bar p \Pi \,.
\ee
These equations of motion for the CDM component are coupled to the other fluid components (baryons, photons and neutrinos) gravitationally via the metric perturbations.

By contrast, if we linearise the equations solved by Newtonian 
simulations we obtain
\begin{subequations}\label{eq:Newton}
\begin{align}  
	 k^2 \Phi^{\rm N} &= 4\pi G a^2 \bar{\rho}_{\rm{cdm}}^{\rm N}  \delta_{\rm cdm}^{\rm N} \,, \\
	 \dot{\delta}_{\rm cdm}^{\rm N} + k v_{\rm cdm}^{\rm N} &= 0 \,, \\ 
   \left[ \partial_\tau + {\cal H} \right] v_{\rm cdm}^{\rm N} &= -k \Phi^{\rm N} \,,
\end{align} 
\end{subequations}
where the superscript N denotes the perturbations of a Newtonian fluid in (unperturbed) flat space.

If we can find a gauge in which the linearised relativistic equations~(\ref{eq:EEfull})--(\ref{eq:NMvelo}) reproduce the linearised Newtonian fluid equations~(\ref{eq:Newton}), then we can use Newtonian simulations to solve for the fully relativistic evolution up to first order. We will now show that the $N$-body \mbox{gauge \cite{Fidler:2015npa}} has this property provided that radiation can be neglected.

\subsection{The \texorpdfstring{$N$}{N}-body gauge}\label{sec:nbodygauge}

The $N$-body gauge 
is defined by: (i) the temporal gauge fixing $ B^{\rm Nb}=  v^{\rm Nb}$, such that the constant-time hypersurfaces are orthogonal to the total matter and radiation 4-velocity, and (ii) the spatial gauge condition $H^\text{Nb}_{\rm L} = 0$, such that the physical volume of a three-dimensional volume element, $\dd^3x$, is unperturbed. 
From equation (\ref{def:curvature}) we see that the spatial gauge condition is equivalent to relating the remaining spatial metric potential to the comoving curvature perturbation, $H_{\rm T}=3\zeta$, which uniquely specifies the spatial gauge independent of the temporal gauge condition.

This spatial gauge condition is crucial, as it implies that the density which a Newtonian simulation computes by a na\"ive counting of particles in a given coordinate volume coincides with the relativistic density, since the physical volume is not perturbed by the relativistic volume deformation, $H_{\rm L}$.\footnote{Other gauges with Newtonian equations of motion at linear order in matter domination exist  \cite{Flender:2012nq,Rampf:2013dxa}, for example the total matter gauge (see appendix~\ref{app:tom}). However, in the presence of a non-vanishing volume deformation ($H_{\rm L}\neq0$), it is not possible to set consistent initial conditions for the simulation, reproducing both the relativistic density and the relativistic positions.} 

Applying the gauge conditions to the Einstein equations~(\ref{eq:EE2})--(\ref{eq:Elast}), we have 
\begin{subequations}
 \begin{align}
 	k^2 \Phi  &= 4 \pi G a^2 \bar \rho \deltaNB \,, \label{eq:PoissonN-bodyFourier} \\ 
 	k^2 \left( A^\text{Nb} + \frac 1 3 H^\text{Nb}_{\rm T} \right) - \left[ \partial_\tau + 2{\cal H}\right] \left( \dot{H}^\text{Nb}_{\rm T}-k B^\text{Nb} \right) &= - 8 \pi G a^2 \bar p \Pi \,, \label{eq:MO} \\
 	{\cal H} A^\text{Nb} - \frac{1}{3} \dot{ H}^\text{Nb}_{\rm{T}} &= 0 \,,  \label{eq:nbA} \\
	\left( \bar \rho + \bar p \right)  A^\text{Nb} &=  -\delta p^\text{Nb} + \frac 2 3 \bar p \Pi  \,. \label{eq:nbA2Pi}
 \end{align}
 \end{subequations}
Equation~(\ref{eq:PoissonN-bodyFourier}) takes the form of the Newtonian Poisson equation, while eq.~(\ref{eq:nbA2Pi}) relates the time perturbation $A^\text{Nb}$ ($\equiv$ the lapse) to the anisotropic stress and pressure perturbations. 
The fluid equations for the CDM component (\ref{eq:NMconti})--(\ref{eq:NMvelo}) in the $N$-body gauge are \cite{Fidler:2015npa}
\begin{subequations}
\begin{align}
    \dot{ \delta}^\text{Nb}_{\rm cdm} + k v^\text{Nb}_{\rm cdm} &= 0 \,, \label{eq:nbodycont} \\
   \left[ \partial_\tau + {\cal H} \right]  v^\text{Nb}_{\rm cdm} &= -k \left( \Phi +  \gammaNB \right) \,. \label{eq:EulerN-bodyFourier}
\end{align}
\end{subequations}
The continuity equation (\ref{eq:nbodycont}) is identical to the Newtonian continuity equation (\ref{eq:NMconti}), since the trace of the spatial metric perturbation (i.e., the volume deformation) is zero in the $N$-body gauge ($H_{\rm L}=0$).
Equation~(\ref{eq:EulerN-bodyFourier}) is the relativistic Euler equation in the $N$-body gauge; it relates the acceleration of the CDM particles to the gradient of the combined potential $\Phi +  \gammaNB$, where the Bardeen potential, $\Phi$, obeys the relativistic Poisson equation (\ref{eq:PoissonN-bodyFourier}) and $ \gammaNB$ is the additional relativistic correction (\ref{def:gammaFourier}) in the $N$-body gauge.

When the energy density of radiation is negligible, the Newtonian potential $\Phi^{\rm N}$, based only on the cold matter, agrees with the Bardeen potential $\Phi$. In this limit, the total pressure and anisotropic stress are small as well, and from eq.\,(\ref{eq:nbA2Pi}) we see directly that the time perturbation $A$ vanishes. Equation~(\ref{eq:nbA}) then implies that we also have $\dot{ H}^\text{Nb}_{\rm T} =0$.
Therefore $\gammaNB$ vanishes since
it is a linear combination (\ref{def:gammaFourier}) of the anisotropic stress $\Pi$ and $\dot{ H}^\text{Nb}_{\rm{T}}$.
The relativistic fluid equations (\ref{eq:PoissonN-bodyFourier})--(\ref{eq:EulerN-bodyFourier}) thus coincide with the Newtonian ones~(\ref{eq:Newton}), where the $N$-body gauge density $\deltaNB_{\rm{cdm}}$, velocity $v^\text{Nb}_{\rm{cdm}}$ and the potential $\Phi$ are all identical to their Newtonian counterparts.

The above analysis shows that the relativistic CDM evolution is correctly described to linear order in the Newtonian simulation if we begin the simulation at a sufficiently late times such that radiation density and pressure are negligible. The particle positions computed in the Newtonian simulation are 
at the correct relativistic coordinates; however the underlying space-time is not Newtonian flat space, but instead it is a non-trivial, dynamical space-time described by the $N$-body gauge metric potentials. As a consequence one should compute observables, such as lensing, in the relativistic space-time using quantities defined in the $N$-body gauge.

\subsection{Newtonian motion gauges}\label{sec:NMgauge}

In the case of non-vanishing radiation, relativistic trajectories in the $N$-body gauge no longer correspond to the Newtonian trajectories. 
To address this, we can instead use the spatial gauge freedom to simply \emph{define} a new relativistic gauge where the particle trajectories $\fett{\chi}_q$ ($\dd\fett{\chi}_q/\dd\tau \equiv \fett{v}_{\rm cdm}$) do match the Newtonian motion:
\be
\label{def:Nmtrajectories}
\fett{\chi}_q^{\rm Nm} (\tau) = \fett{\chi}_q^{\rm N} (\tau) \,.
\ee
We call this a {Newtonian motion} gauge (denoted by Nm). 
The spatial coordinates used in the Newtonian simulation can then be identified with the coordinates in this gauge even in the presence of radiation:
\be
\label{def:Nm}
\fett{x}^{\rm Nm} = \fett{x}^{\rm N} \,.
\ee
The relativistic particle positions are understood to be not on a flat Newtonian space, but on the non-trivial space-time of the Newtonian motion gauge. We will now show how the spatial gauge condition  (cf.\,eq.\,(\ref{def:Nmtrajectories})), matching the relativistic trajectories to the Newtonian ones, can be realised explicitly. 

For the relativistic Euler equation (\ref{eq:NMvelo}) to reproduce the Newtonian trajectories we must identify the combined relativistic potential $\Phi + \gamma$ with the Newtonian potential, $\Phi^{\rm N}$. We thus realise the Newtonian motion gauge by requiring that 
\be \label{eq:NMgauge}
\gamma^{\rm Nm} = \Phi^{\rm N} - \Phi\, ,
\ee
where the Newtonian simulations compute $\Phi^{\rm N}$ using the Newtonian Poisson equation,
\be
\label{def:NPhi}
 k^2 \Phi^{\rm N} = 
   4\pi G a^2 \bar \rho_{\rm cdm} \delta_{\rm cdm}^{\rm N} \,.
\ee
Note that the Newtonian CDM density, $\delta_{\rm cdm}^{\rm N}$, is the \emph{coordinate} CDM density, i.e., the physical density plus the GR volume deformation in the Newtonian motion gauge:
\begin{equation}
\label{def:Ndensity}
\delta_{\rm cdm}^{\rm N} \equiv \delta^\text{Nm}_{\rm cdm} + 3 H_{\rm L}^\text{Nm} \,.
\end{equation}
In this combination, the temporal gauge dependence cancels (see eqs.\,(\ref{eq:generalgaugetrafoHL}) and~(\ref{eq:generalgaugetraforho})), leaving the Newtonian 
density dependent only on the spatial gauge transformation, 
\begin{equation}
\label{deltaNtrafo}
\delta_{\rm cdm}^{\rm N} = \tilde{\delta}^{\rm N}_{\rm cdm} - kL \,.
\end{equation}
It obeys the continuity equation, 
\begin{equation}
\label{eq:vNm}
\dot \delta_{\rm cdm}^{\rm N} + k v^\text{Nm}_{\rm cdm} = 0,
\end{equation}
while the relativistic Euler equation in the Newtonian motion gauge reads,
\be \label{eq:NMveloexp}
 \left[ \partial_\tau + {\cal H} \right] v^{\rm Nm}_{\rm cdm} = -k  \Phi^{\rm N} \,.
\ee
The set of equations~(\ref{def:NPhi}),~(\ref{eq:vNm}) and~(\ref{eq:NMveloexp}) coincide with the linearised Newtonian equations (\ref{eq:Newton}).
As a consequence, Newtonian  simulations can be used to compute the relativistic velocities in the Newtonian motion gauge and thus update the CDM positions consistent with GR up to first order. 

While the velocities (and positions) in the Newtonian simulation directly match the relativistic velocities (and coordinates), the density and potential in the Newtonian simulation do not correspond to their relativistic counterparts.
From the Newtonian motion gauge viewpoint, the Newtonian density (\ref{def:Ndensity}) and potential (\ref{def:NPhi}) are only auxiliary quantities that are used in the simulation to move the particles according to GR. However, these Newtonian quantities can be related to the relativistic density and potential at any time through equations (\ref{def:Ndensity}) and (\ref{eq:NMgauge}).

\section{Choosing a Newtonian-motion gauge}\label{NMgauge}

The Newtonian motion definition~(\ref{def:Nmtrajectories}) does not uniquely specify the gauge, and so there are many possible Newtonian-motion gauge choices. In this section we will discuss the residual freedom in the specification of the Newtonian-motion gauges, and present one useful example of a Newtonian-motion gauge choice that is closely related to the $N$-body gauge.

Matching the Newtonian trajectories requires only satisfying the condition~(\ref{eq:NMgauge}), where the Bardeen potential is gauge invariant and $\Phi^{\rm N}$ depends only on the spatial gauge. Similarly, $\gamma^{\rm Nm}$ is only sensitive to the spatial gauge choice, $L$ in (\ref{def:generalgaugetrafo}), since it is computed from the metric potential $H_{\rm T}$ and the anisotropic stress $\Pi$ (see eqs.~(\ref{eq:generalgaugetrafoHT}),~(\ref{eq:generalgaugetrafoPi})). As a consequence, we have complete freedom to choose any temporal gauge while still satisfying this Newtonian motion condition. 

Furthermore,  the condition~(\ref{eq:NMgauge}) is equivalent to a second-order differential equation for $L$, and two additional boundary conditions are required to specify a unique spatial gauge. The Newtonian motion gauges thus have a residual spatial gauge freedom that can be used to specify any desired initial Newtonian density, by choosing an appropriate initial value of $L$ [see equation (\ref{deltaNtrafo})], and any desired initial velocity, by choosing the initial value for $\dot{L}$ [see equation~(\ref{eq:generalgaugetrafov})]. 

We can construct a Newtonian motion gauge for any possible choice of initial conditions used in the simulation. This is not surprising, as the condition~(\ref{eq:NMgauge}) forces particles to follow Newtonian trajectories, but it does not assume anything about their initial positions and velocities.

We should use this freedom to ensure that the small-scale evolution in the given Newtonian motion gauge remains as close as possible to the $N$-body gauge, which we have argued provides a good description of the matter evolution in the absence of radiation (see also appendix~\ref{app:A}).
Thus, we assume the relativistic corrections only affect the linear evolution, i.e., above the non-linear scale, and hence can be described by the linear Einstein--Boltzmann code. 
In section~\ref{sec:nbodyNM} we will provide a choice of gauge with this property.

In summary there exists a broad class of Newtonian motion gauges, specified by both the temporal gauge condition and the choice of initial spatial gauge conditions. In appendix~\ref{app:A}, we discuss how the choice of the initial conditions can help to ensure that the metric potentials stay small.  The temporal gauge is chosen for computational convenience, and below we make a simple choice which reduces to the $N$-body gauge in the absence of radiation. However, other choices are also possible, and in appendix~\ref{app:Poisson} we explore an example that corresponds to the temporal slicing used in the Poisson gauge.

\subsection{A simple Newtonian motion gauge}\label{sec:nbodyNM}

First, we fix the temporal gauge by requiring that the constant time hypersurfaces are orthogonal to worldlines comoving with the total matter, i.e.,  $v^{\rm Nm}=B^{\rm Nm}$; we are thus defining a comoving-orthogonal Newtonian motion gauge. 
This comoving-orthogonal temporal gauge condition is shared by the $N$-body gauge, defined in section~\ref{sec:nbodygauge}, as well as the total matter gauge and (in the absence of radiation) the synchronous-comoving gauge~\cite{Malik:2008im}.

Second, we fix the remaining spatial gauge freedom by identifying the spatial gauge at the initial time with the $N$-body gauge; we require $H_{\rm{L}}(\tau_{\rm ini}) = 0$ and $\dot{H}_{\rm{L}}(\tau_{\rm ini}) = 0$.

We have seen that in the absence of radiation the Newtonian simulation correctly reproduces the relativistic solution to first order in the $N$-body gauge, i.e., in that limit the $N$-body gauge {\em is} a Newtonian motion gauge with $H_{\rm L}=0$ and $H_{\rm T}=3\zeta=$~constant.
More generally, a Newtonian simulation using initial dark matter displacements and velocities in the $N$-body gauge computes the dark matter positions in the Newtonian motion gauge that {\em initially} coincides with the $N$-body gauge space-time, but in the presence of radiation the Newtonian motion gauge departs from the $N$-body gauge ($H_{\rm L}\neq0$, ${H}_{\rm T}\neq3\zeta$). Thus the Newtonian motion gauge accommodates all relativistic corrections, including radiation, through the time-dependence of the metric potentials up to first order, while Newtonian simulations solve for the gravitational collapse of matter on small-scales in the fully non-linear but Newtonian theory.

\begin{figure}[t]
	\includegraphics[width=\textwidth]{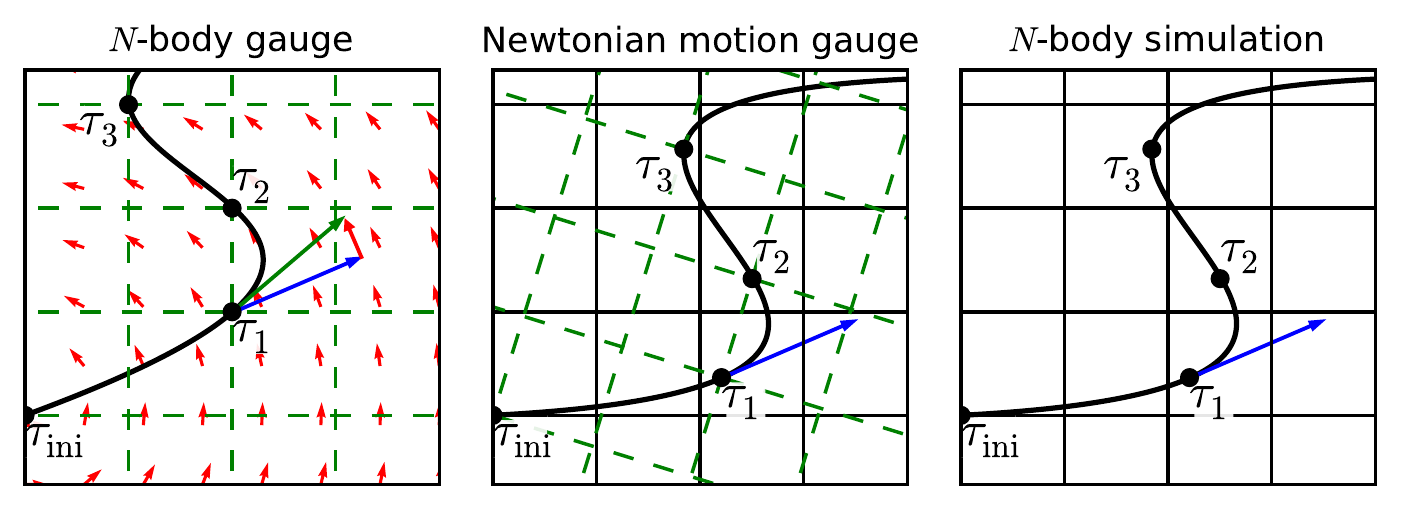}
	\caption{Sketch of a particle trajectory (black solid line) in three different coordinate systems/gauges. For illustration purposes, the impact of radiation is assumed here to be just a time-independent rotation, and this is denoted with red arrows. In the $N$-body gauge particles move according to their relativistic velocity (green arrow), which is given by a Newtonian component (blue arrow) and a correction due to radiation (red arrow).
In the Newtonian-motion (Nm) gauge, by contrast, particles are at the same initial position as in the $N$-body gauge at $\tau_{\rm ini}$, but then only move according to the Newtonian velocity. Particles thus behave as they would travel on pure Newtonian trajectories, whereas the impact of radiation is embedded in the Nm coordinates. Also note that the particle position in the simulation matches the particle position in the Nm gauge at any time. 
Combining the output of a Newtonian simulation with the relativistic space-time of the Newtonian motion gauge results in a consistent relativistic solution including radiation.
}
 \label{fig:NMD}
\end{figure}
This idea is sketched in figure~\ref{fig:NMD}. In the $N$-body gauge, the presence of radiation causes the CDM particles to depart from the Newtonian trajectory, while in the Newtonian motion gauge the CDM particles follow the Newtonian trajectories by construction. In the Newtonian motion gauge, the impact of radiation is incorporated in the evolution of the space-time, with the metric being deformed by the presence of radiation compared with the $N$-body gauge metric.  


\section{Application of the Newtonian motion gauge to Newtonian simulations}\label{sec:appN-bodySim}

In this section we explicitly provide the steps needed to incorporate radiation and GR to linear order in a Newtonian
simulation. 
After discussing the initial conditions for the simulations, we describe two methods of incorporating the relativistic corrections after the simulations have been run.

\subsection{Setting the initial conditions}
\label{sec:initialconditions}

By using our simple Newtonian motion gauge,  described in section~\ref{sec:nbodyNM}, the initial density, displacement and velocity fields can be specified in the $N$-body gauge, where the relativistic and Newtonian matter densities coincide. 
The matter power spectra in the $N$-body gauge can be computed at first order using Einstein--Boltzmann codes (such as \CLASS{} \cite{Blas:2011rf}). In a given simulation one must then select random amplitudes and phases for each Fourier mode. 

The initial displacement field for CDM and baryons can be obtained from the $N$-body gauge density field using the Zel'dovich approximation for each species, $\vec{\nabla} \cdot \fett{\Psi}_\alpha=-\delta_\alpha$.
The appropriate fraction of the simulation particles should be displaced according to the CDM density field calculated by Einstein--Boltzmann codes, while the rest are displaced according to the baryon density. 
Similarly the initial velocity fields for CDM and baryons should be computed separately using the output from first-order Einstein--Boltzmann codes.

Note that the synchronous-comoving gauge density, often used to generate initial conditions, is almost identical to the $N$-body gauge density, except that the $N$-body gauge is comoving-orthogonal with respect to the total matter and radiation while the standard synchronous gauge is comoving with respect to the CDM. Thus there will be small differences in the temporal gauge, and hence the density fields, due to the finite velocities of baryons and radiation with respect to the CDM, especially for simulations starting at high redshift.

In conventional Newtonian 
simulations, the matter power spectrum is often evaluated at the present time, using first-order Einstein--Boltzmann codes that include the effect of radiation, and then rescaled back to the initial time using the linear Newtonian equations in the absence of radiation. This back-scaling is done in order to ensure that the Newtonian simulations, which do not include radiation, can reproduce the correct linear matter power spectrum at present at the expense of starting from a fictitious initial density field.\footnote{Note that in this approach the CDM and baryons have very similar displacements at the present time and thus are usually treated identically in Newtonian 
simulations, while in reality they have different initial displacements and velocities.}

Using the Newtonian-motion gauge, the back-scaling is no longer needed. All physics (at linear order), including the effect of radiation, is included in the Newtonian-motion gauge space-time. As a consequence the actual density and velocity fields calculated by Einstein--Boltzmann codes at the initial time should be used without back-scaling.

After setting up these initial conditions, the Newtonian  
simulation can be used, without any modification, to compute the particle trajectories in the Newtonian motion gauge. A simulation initialised following this recipe computes the correct relativistic position of the particles, which then need to be interpreted as trajectories in the Newtonian motion space-time in order to obtain a consistent relativistic result. In the following we present two alternative methods for providing such an interpretation.

\subsection{Post-processing of simulations using the Newtonian motion space-time}\label{sec:post-processing}

Once the Newtonian 
simulation has been run, relativistic observables can be computed from the Newtonian  
results where the resulting displacements are embedded in the Newtonian-motion space-time. This curved space-time can then be used to obtain a fully relativistic interpretation of the Newtonian simulation results.

While the Newtonian motion of CDM particles is accurately computed by Newtonian 
simulations in full non-linearity, we are left with the task to solve the underlying relativistic space-time characterised by the metric potentials in the Newtonian motion gauge up to first order. 
A real-space version of the metric perturbations is 
generated from the first order results of an Einstein--Boltzmann code such as \CLASS{}, using the same realisation (amplitudes and phases of each Fourier mode) used in the Newtonian simulation when generating the initial conditions.
Any desired observable can then be constructed from the output of the simulation and from the metric potentials of the Newtonian motion gauge.

The evolution of the relativistic space-time in the presence of radiation is given by the evolution of the metric potentials,
which we compute within the Newtonian motion gauge using a modified version of the Einstein--Boltzmann code \CLASS{}. 
The perturbed metric (\ref{eq:metric}) in Newtonian motion gauge is written as
\begin{subequations}
\begin{align}
  g_{00} &= -a^2 \left( 1 + 2A^\text{Nm} \right) \,, \\
  g_{0i} &=  a^2\, \ii \hat k_i B^\text{Nm} \,, \\
  g_{ij} &= a^2 \left[ \delta_{ij} \left( 1 + 2 H^\text{Nm}_{\rm L} \right) + 2 \left( \delta_{ij}/3 - \hat k_i \hat k_j \right) H^\text{Nm}_{\rm T} \right] \,.
\end{align}
\end{subequations}
Let us discuss the perturbations within the Einstein equations for each of the components in the Newtonian motion gauge metric defined in section~\ref{sec:nbodyNM}: 
\begin{itemize}
\item The lapse function perturbation $A^\text{Nm}$ is determined by the relativistic stress according to the Einstein equation (\ref{eq:Elast}) in a comoving-orthogonal gauge
\be
	\left( \bar \rho + \bar p \right) A^\text{Nm} = \frac 2 3 \bar p \Pi - \delta p^\text{Nm}  \,.  \label{last2}
\ee
The lapse perturbation is directly related to the anisotropic stress and grows during radiation domination, whereas it decays in the matter- and $\Lambda$-dominated eras. 
Note that $A^{\rm Nm}$ only depends on the temporal gauge and is thus identical to the $N$-body gauge lapse function, $A^{\rm Nm}=A^{\rm Nb}$ (cf.\,eq.\,(\ref{eq:nbA2Pi})).
\item The shift perturbation is given by the total velocity via the comoving-orthogonal temporal gauge condition 
\be
B^{\rm Nm}=v^{\rm Nm}\,.
\ee
\item The spatial gauge condition of the Newtonian motion gauge, eq.\,(\ref{eq:NMgauge}), together with the Einstein equation~(\ref{eq:EEA}), specifies $H_{\rm L}$, 
\be
 -4 \pi G a^2 \bar{\rho}_{\rm cdm} (3 H^{\rm Nm}_{\rm L} + \delta^{\rm Nm}_{\rm cdm}) = k^2 A^{\rm Nm} + \left( \partial_\tau + \mathcal{H} \right) k B^{\rm Nm} \,.
\ee
The volume deformation, $H^{\rm Nm}_{\rm L}$, is initially vanishing and then grows due to the presence of radiation. In addition once the volume deformation is non-zero, it starts to evolve as the Universe expands, keeping track of the gravitational growth of the perturbations induced from the early radiation perturbations.
\item The time derivative of the spatial potential $H_{\rm T}$ is provided by the Einstein \mbox{equation~(\ref{eq:EEB})} in a comoving-orthogonal gauge
\be
 \frac 1 3 \dot{H}^\text{Nm}_{\rm T} = {\cal H} A^\text{Nm} - \dot{H}^\text{Nm} _{\rm L} \,.  \label{last1}
\ee
\end{itemize}

\begin{figure}[t]
	\begin{centering}
	\includegraphics[width=1.0\textwidth]{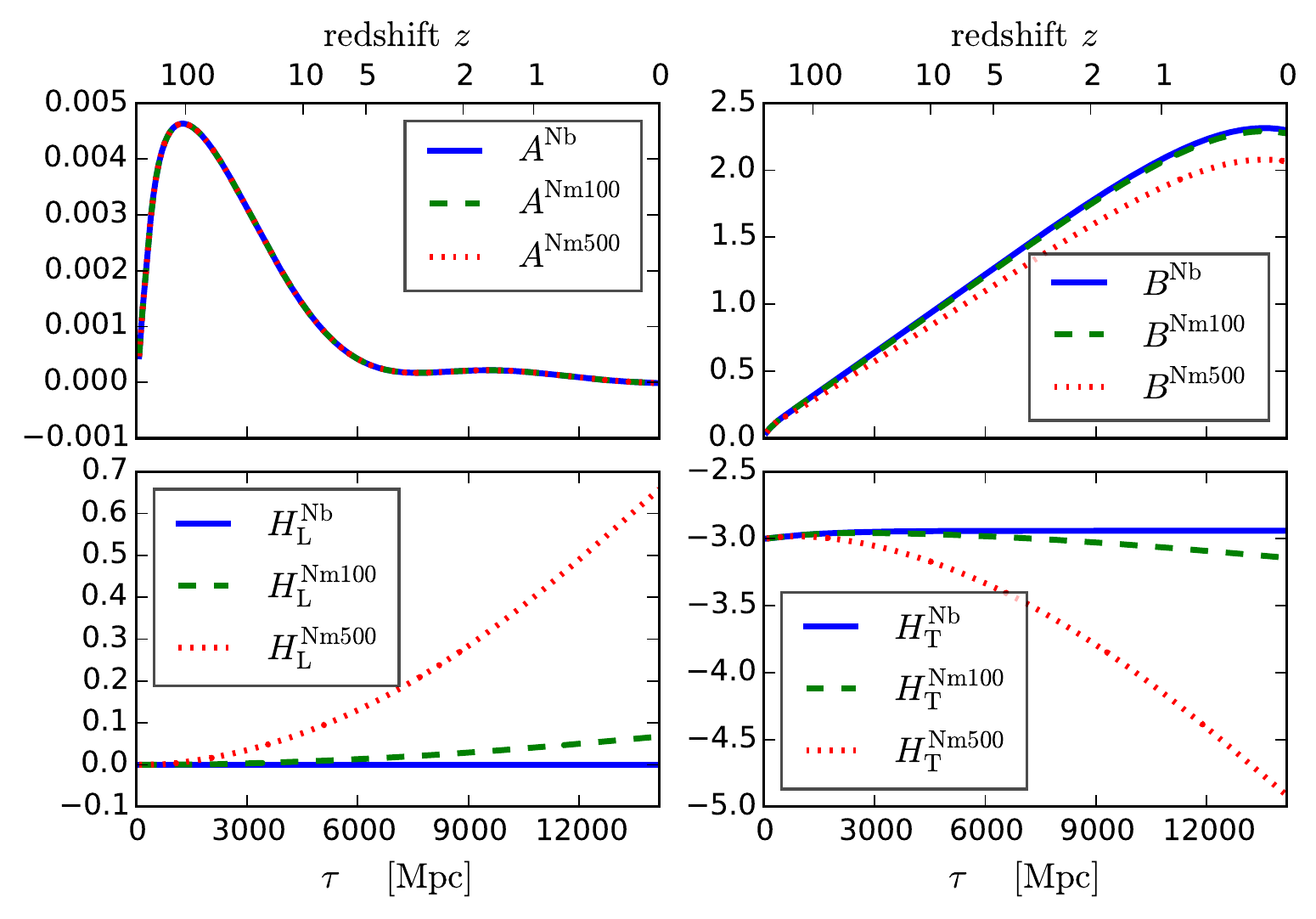}
	\caption{Time evolution of the metric potentials for the wavemode $k=10^{-3}\text{Mpc}^{-1}$ in the $N$-body gauge and two Newtonian motion gauges. Here we use the superscripts Nm100 and Nm500 to denote the Newtonian motion gauge initialised at $z=100$ and $z=500$ respectively. The lapse perturbation $A$ is identical in all three gauges and decays in matter domination. The  $N$-body gauge shift vector $B^\text{Nb}$ is not identical to $B^\text{Nm100}$ and $B^\text{Nm500}$ due to the effect of radiation on the velocities, but the discrepancy is small especially for $B^\text{Nm100}$. In all three gauges $B$ grows slowly with the dark matter velocity. The volume deformation $H_{\rm{L}}$ vanishes in the $N$-body gauge while it grows in the Newtonian motion gauges. In the same way, the spatial potential $H_{\rm{T}}$ in the Newtonian motion gauge departs from the $N$-body gauge value of $3\zeta$. Perturbations are normalised relative to $\zeta=-1$ on super-horizon scales which is the default setting in \CLASS{}.}
\label{fig:metric}
\end{centering}
\end{figure}

The resulting metric potentials
are shown in figure~\ref{fig:metric} for the wavemode $k=0.001$\,Mpc$^{-1}$. Solid lines describe the metric potentials in the $N$-body gauge described in section~\ref{sec:nbodygauge}, while dashed (dotted) lines are results for the Newtonian motion gauge which coincides with the $N$-body gauge when initialised at $z=100$  ($z=500$). 

As expected, the lapse perturbation $A$ (upper left panel) grows and then decays, while the shift $B$ (upper right panel) grows with the dark matter velocities. As evident from the figure, the shift $B$ in the $N$-body and Newtonian motion gauge differs, while the time perturbation $A$ is identical in both gauges.

In the $N$-body gauge the volume deformation $H_{\rm L}$ vanishes by definition. 
As long as the volume deformation is small, the Newtonian motion gauge remains close to the $N$-body gauge.
This can be seen in the lower left panel of figure~\ref{fig:metric} for the Newtonian motion gauge initialised at $z=100$. As long as $H_{\rm{L}}$ is small, the other metric potentials must agree with the $N$-body gauge potentials, and indeed the spatial potential $H_{\rm{T}}$ (lower right panel) is almost identical in both gauges. By contrast, the Newtonian motion gauge initialised at $z=500$ deviates significantly from the $N$-body gauge; $H_{\rm L}$ grows quickly due to the presence of radiation initially. In the matter dominated era, the volume deformation then continues to grow with the dark matter growth function (as we shall show in section~\ref{sec:postprocessNb}). As $H_{\rm L}$ grows, the metric potential $H_{\rm{T}}$ also deviates significantly from the $N$-body gauge value of $3\zeta$.

\begin{figure}[t]
 	\begin{centering}
 	\includegraphics[width=1.0\textwidth]{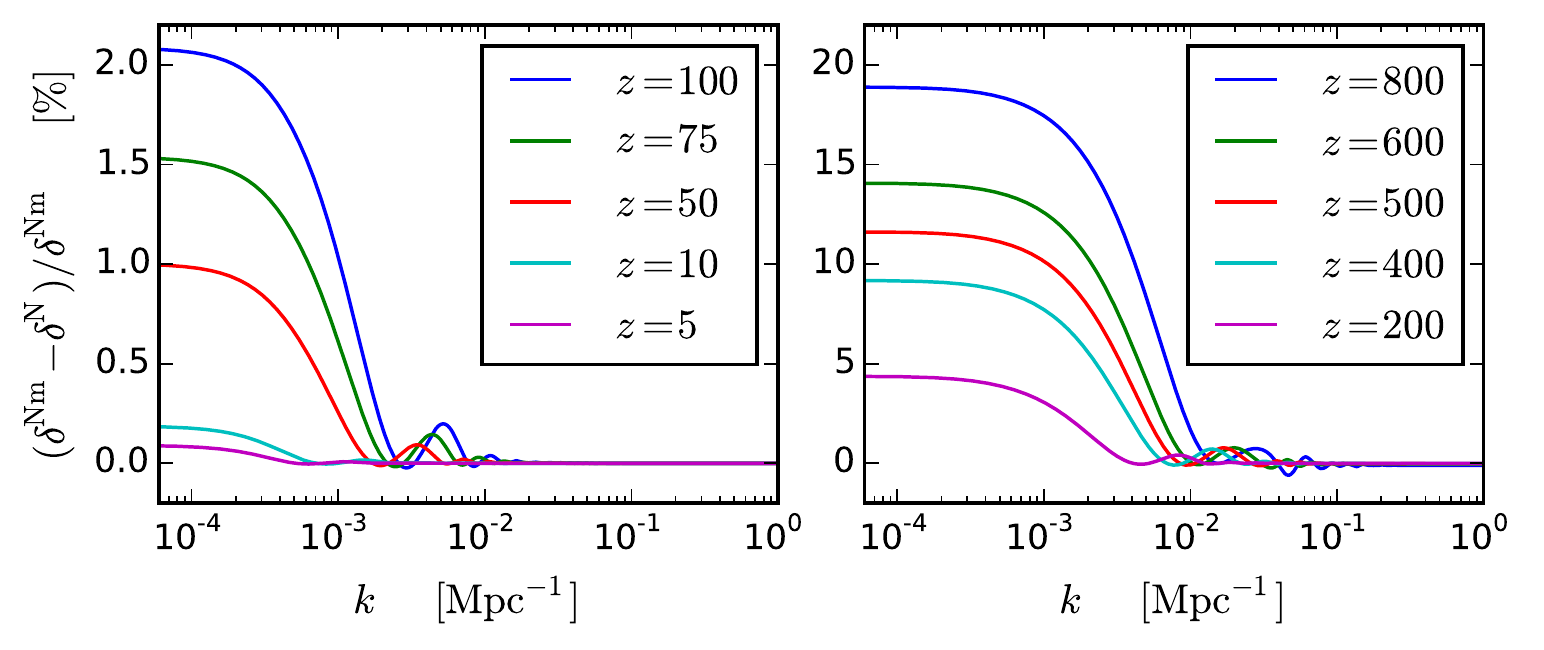}
 	\caption{The relative percentage difference at redshift $z=0$ between the relativistic density $\delta^\text{Nm}=\delta^\text{Nb}$ and the Newtonian density $\delta^{\rm N}$ for a simulation initialised at various redshifts due to the presence of radiation. The difference at large scales is up to 1\% [4\,\%] for typical initial redshifts of $z=50$ [$z=200$].
} \label{fig:correction}
 \end{centering}
 \end{figure}

In figure~\ref{fig:correction} we show the impact of the volume deformation on the linear density field, given by eq.\,(\ref{def:Ndensity}). We compute the linear relativistic matter fluctuation and compare it to the contribution induced from the volume deformation when the Newtonian motion gauge is initialised at various redshifts. The volume deformation due to the presence of radiation is most important on the largest scales, and for an initial time of $z=50$ [$z=200$] it induces 1\% [4\%] corrections to the matter density, with earlier initial times resulting in much bigger corrections.  

\subsection{Post-processing by gauge transformation to \texorpdfstring{$N$}{N}-body gauge}\label{sec:postprocessNb}
 \begin{figure}[t]
 	\includegraphics[width=1.0\textwidth]{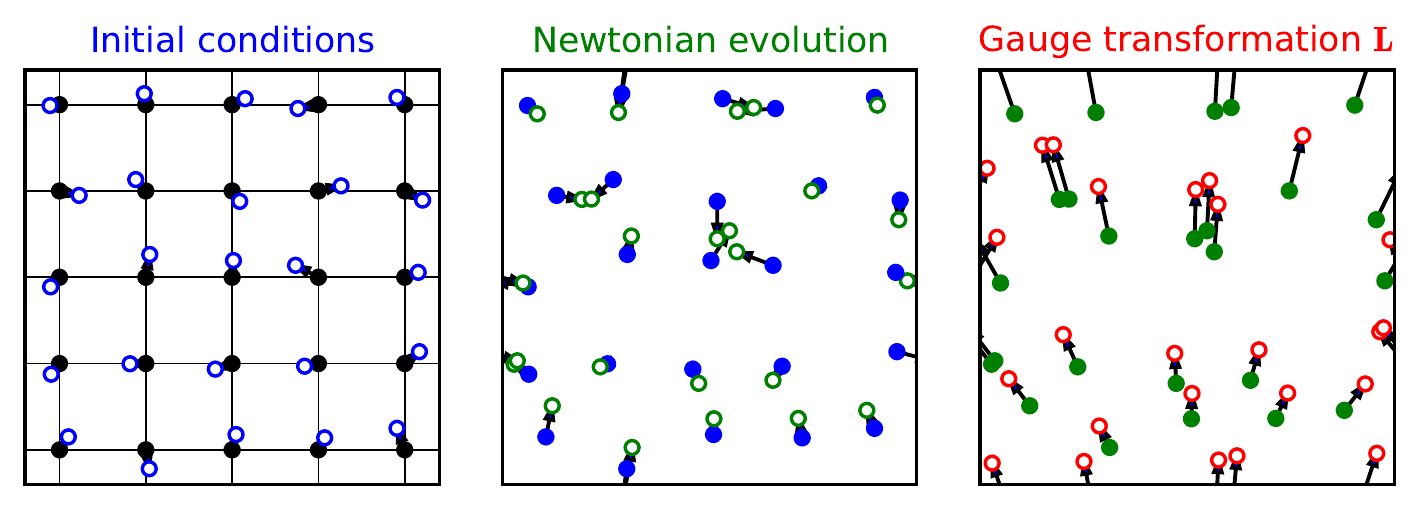}
 	\caption{Sketch of the procedure described in section~\ref{sec:postprocessNb} of post processing by gauge transformation. The $N$-body gauge and the Nm gauge coincides when initial conditions are imposed. After some Newtonian evolution this is no longer the case however, but we can use the spatial gauge transformation $\fett{L}$ to go back to $N$-body gauge. In practice this would be achieved by displacing all particles by $\fett{L}$ in each snapshot.} \label{fig:cartoon}
 \end{figure}
Instead of performing the analysis of the simulation results directly in the Newtonian motion gauge, it is possible to use a gauge transformation from the Newtonian motion gauge to obtain relativistic results in any other well-defined gauge. As an example we will compute the gauge transformation connecting the Newtonian motion gauge quantities with the $N$-body gauge defined in section~\ref{sec:nbodygauge}. 

As the temporal gauge condition in the Newtonian motion and $N$-body gauge are identical \mbox{($\tau^{\rm{Nb}} = \tau^{\rm{Nm}}$),} the gauge transformation is purely spatial. We define 
\be
\label{def:Nmb}
  \fett{x}^{\rm{Nb}} = \fett{x}^{\rm{Nm}} + \fett{L}^\Nmb \,, \qquad \fett{L}^\Nmb = -k^{-1} \nabla L^\Nmb \,,
\ee 
where
\begin{equation}
	\dot{L}^\Nmb = v^\text{Nb}_{\rm{cdm}} - v^\text{Nm}_{\rm{cdm}} \,.
\end{equation}
In particular the coordinate transformation~(\ref{def:Nmb}) can be applied directly to the particle displacements from a Newtonian simulation (since these coincide with the relativistic displacements in the Newtonian-motion gauge) in order to obtain the relativistic particle displacements in the $N$-body gauge. 

An outline of this procedure is shown in figure~\ref{fig:cartoon}. 
We obtain a real-space version of the gauge transformation $\fett{L}^\Nmb(x)$ using the particular realisation (amplitudes and phases of each Fourier mode) employed to generate the particular initial conditions used in the Newtonian 
simulation.

Both $L^\Nmb$ and its time derivative vanish initially since our simple Newtonian-motion gauge coincides with the $N$-body gauge at the initial time by construction. 
The second derivative of the gauge transformation is given by comparing equations~(\ref{eq:EulerN-bodyFourier}) and~(\ref{eq:NMveloexp}) which give the difference of the forces acting on the dark matter in the two gauges:
\begin{equation} \label{eq:gaugein1}
	  \ddot{L}^\Nmb + {\cal H} \dot{L}^\Nmb =  - k \left(\Phi-\Phi^{\rm{N}} + \gammaNB \right) \,,
\end{equation}
where $\gammaNB$ is the $N$-body gauge relativistic correction given in equation~(\ref{def:gammaFourier}).
The Bardeen potential, $\Phi$, is sourced by the density of all species, while the Newtonian potential, $\Phi^{\rm N}$, only includes the non-relativistic matter densities as they appear in the simulation. 
From equations~(\ref{eq:PoissonN-bodyFourier}) and~(\ref{def:NPhi}) we have
\begin{equation}
 \label{eq:thistoo}
	 k^2\left( \Phi - \Phi^{\rm N} \right) = 4\pi G a^2 \left( \bar{\rho}\delta^{\rm{Nb}} - \bar{\rho}_{\rm{cdm}}\delta^{\rm{N}}\right)  \,.
\end{equation}

For convenience we split $\bar{\rho}\delta^{\rm{Nb}} = \bar{\rho}_{\rm{cdm}}\delta^\text{Nb}_{\rm{cdm}}+\bar{\rho}_{\rm{other}}\delta^\text{Nb}_{\rm{other}}$, with $\delta^\text{Nb}_{\rm{other}}$ including the relativistic species not present in a Newtonian simulation. The difference between the dark matter density in the $N$-body gauge and the Newtonian simulation is given by the volume deformation in the Newtonian motion gauge 
\be \label{eq:diffdens}
\delta^\text{Nb}_{\rm{cdm}} - \delta^\text{N} = \delta^\text{Nb}_{\rm{cdm}} - \delta^\text{Nm}_{\rm{cdm}} -3 H^{\rm Nm}_{\rm L} = -3 H^{\rm Nm}_{\rm L} \,.
\ee 
We employ the gauge transformation of the volume deformation~(\ref{eq:generalgaugetrafoHL}) and since $H_{\rm L}^{\rm Nb}$ is vanishing we find:
\be
k L^\Nmb = 3 H^{\rm Nm}_{\rm L}\,.
\ee
This relation directly specifies the gauge transformation, given that we know the Newtonian motion gauge volume deformation. However, here we intend to compute the gauge transformation using only perturbations in the $N$-body gauge, suitable for a numerical implementation in a Boltzmann code running in the $N$-body gauge. We insert the relation into eq.~(\ref{eq:diffdens}) and find:
\be
 \label{eq:deltaNbcdm}
 \delta^\text{Nb}_{\rm{cdm}} - \delta^{\rm{N}} = - k L^\Nmb \,.
\ee
This relation simply states that, since there is no volume deformation in the $N$-body gauge, the relativistic matter density in the $N$-body gauge is obtained by displacing the particles with respect to the Newtonian simulation according to $L^\Nmb$, see figure~\ref{fig:cartoon}. In this way, the $N$-body gauge displacements track the relativistic effects of radiation which are absent in the Newtonian simulation. 

Finally, substituting equations~(\ref{eq:thistoo}) and~(\ref{eq:deltaNbcdm})
into the equation for the gauge transformation (\ref{eq:gaugein1}) we obtain
\begin{align}
	  \ddot{L}^\Nmb &+ {\cal H} \dot{L}^\Nmb -  4\pi G a^2 \bar{\rho}_{\rm{cdm}} L^\Nmb= - k \gammaNB
	  - 4\pi G a^2 k^{-1} \bar{\rho}_{\rm{other}} \delta^\text{Nb}_{\rm{other}} \,. \label{ODE-L}
\end{align}
The first source term $\gammaNB$ is the relativistic force acting on the dark matter particles in the $N$-body gauge and the second source term 
$\sim \delta^\text{Nb}_{\rm{other}}$ is due to the presence of the other relativistic species, absent in Newtonian simulations, contributing to the full gravitational potential. 
Both cause CDM trajectories in the $N$-body gauge to deviate from the Newtonian motion.  

The homogenous solution of~(\ref{ODE-L}) is the linear dark matter growth function. Once a particle in the $N$-body gauge is displaced from the Newtonian position, this corresponds to a difference in the density perturbation and hence the trajectories in the simulation and the $N$-body gauge will further deviate
as this density perturbation grows, even after the radiation becomes negligible. 

 \begin{figure}[t]
 	\begin{centering}
	\includegraphics[width=1.0\textwidth]{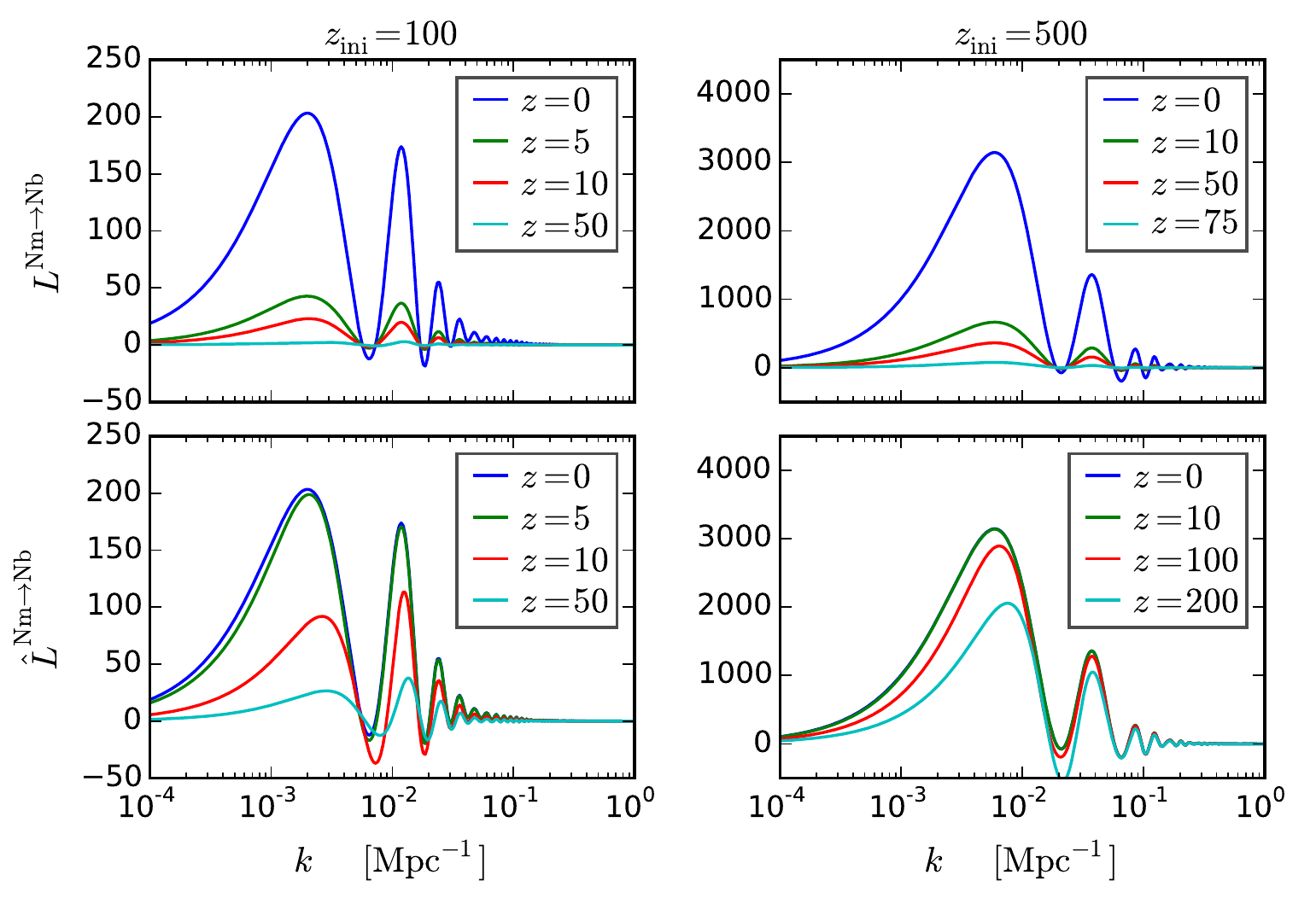}
 	\caption{The top row shows the gauge transformation $L^\Nmb$ defined in equation~(\ref{def:Nmb}) (in Fourier space) connecting the Newtonian motion gauge to the $N$-body gauge as a function of $k$ for two initial redshifts $z_\text{ini}=100$ and $z_\text{ini}=500$. $L^\Nmb$ grows partly due to radiation sources and partly due to density growth as explained in the text. In the bottom row we have shown $\hat{L}^\Nmb$ which is just $L^\Nmb$ normalised by the linear dark matter growth function. This isolates the effect of the source term, and we can see that it vanishes at the late times as expected. See the caption of figure~\ref{fig:metric} for the normalisation of perturbations.}\label{fig:timee}
 \end{centering}
 \end{figure}

We compute the gauge generator $L$ in Fourier space again using \CLASS{}.\footnote{The modified version of \CLASS{} which solves $L^\Nmb$ and the $N$-body gauge metric potentials is available upon request: \href{mailto:thomas.tram@port.ac.uk}{thomas.tram@port.ac.uk}.}
The time evolution of the gauge transformation is explicitly shown in figure~\ref{fig:timee}, presenting both $L^\Nmb$ and $L^\Nmb$ divided by the linear matter growth function. After $z=10$ the latter curves overlap showing that radiation source terms becomes negligible. Before $z=10$ radiation actively drives the gauge transformation and causes the coordinate systems of the $N$-body gauge and the Newtonian motion gauge to deviate.
 \begin{figure}[t]
 	\begin{centering}
 	\includegraphics[width=1.0\textwidth]{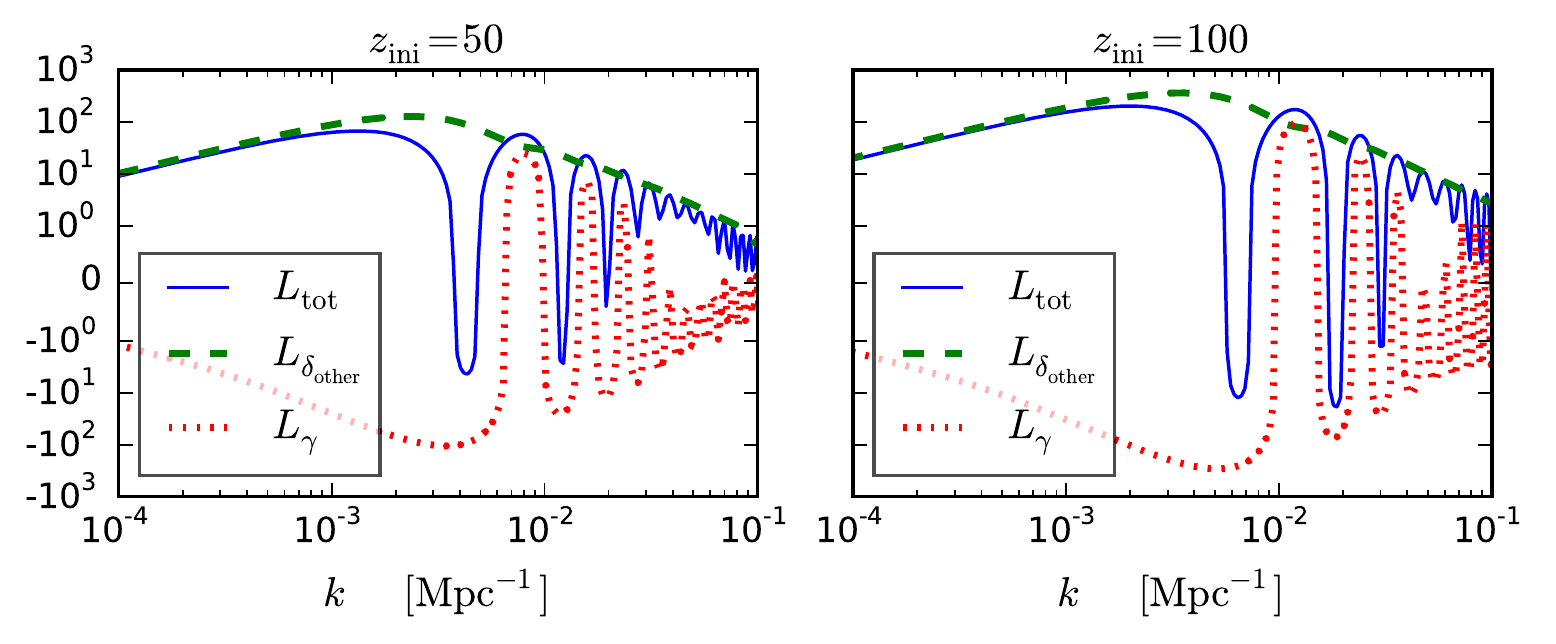}
 	\caption{The gauge transformation $L$ at redshift $z=0$ initialised at $z=50$ (left panel) and $z=100$ (right panel). The gauge transformation is sourced by the relativistic force $\gamma$ as well as density contributions from relativistic species, $\delta_\text{other}$.  The contribution from $\delta_\text{other}$ (dashed lines) is smooth and positive while the $\gamma$ contribution (dotted lines) is oscillatory and mostly negative (i.e., anti-correlated with $\Phi(\fett{k})$). See the caption of figure~\ref{fig:metric} for the normalisation of perturbations.} \label{fig:compare}
 \end{centering}
 \end{figure}

Figure~\ref{fig:compare} shows the gauge transformation induced from the relativistic force $\gamma$ and the impact of the radiation density on the Bardeen potential separately. The total gauge transformation at the linear order is the sum of both. We see that both contributions have a similar scale dependence and comparable size. The term induced by $\gamma$ is oscillatory, being related to the anisotropic stress, while the sources related to the radiation density are significantly smoother.

\section{Conclusions}\label{sec:conclusions}

General relativistic effects in structure formation have to be taken into account for the next generation of large-scale structure surveys such as the SKA\footnote{\tt www.skatelescope.org} and the EUCLID mission.\footnote{\tt sci.esa.int/euclid} Instead of modifying existing Newtonian simulations, we provide the relativistic solution by computing the space-time on which these simulations should be interpreted.  We have defined a class of gauges, called Newtonian motion gauges, that by construction keep the matter trajectories in Newtonian simulations unchanged. 

To maintain Newtonian-like trajectories within a general relativistic multi-fluid Universe, we impose a specific spatial gauge condition, eq.\,(\ref{eq:NMgauge}), which all Newtonian motion gauges must obey. This is a spatial gauge condition which ensures that the gravitational acceleration of the CDM particles is sourced by a Newtonian-like Poisson equation, itself solved by taking only the gravitational interaction of CDM into account.
This, together with the identification of a Newtonian density contrast (\ref{def:Ndensity}), produces a closed sub-system of Newtonian CDM fluid equations, whereas the evolution of the relativistic fluid species is governed by the Einstein-Boltzmann equations on the relativistic space-time. 

The Newtonian motion gauges introduce a convenient split between the computation of the particle trajectories in the Newtonian theory and the evolution of the metric potentials in the relativistic theory. The highly non-linear collapse of matter can be simulated in full non-linearity in Newtonian simulations. For this task the existing and optimised Newtonian codes can be used without modifications. 
On the other hand, the underlying metric is described by linear perturbation theory and can be obtained by modifying conventional Einstein--Boltzmann codes. The Newtonian and relativistic problem is solved independently, with the Newtonian-motion gauge providing the framework to combine the two when generating initial conditions and when interpreting the output 
(see section~\ref{sec:initialconditions}).
%
%
Since a linear Einstein--Boltzmann code requires only a fraction of the time needed for a typical Newtonian simulation, the relativistic evolution can be computed at almost no extra cost.   
A modified version of \CLASS{} which computes the evolution of the relativistic space-time is available upon request. 

We have shown that the present-day matter density calculated in Newtonian simulations receives corrections of up to 1\% [4\%] on large scales in GR due to the effect of radiation when initialised at redshifts of $z=50$ [$z=200$]. This is accounted for in conventional $N$-body simulations by introducing a fictitious correction to the initial matter density at high redshifts such that the present-day matter power spectrum reproduces the power spectrum calculated by Einstein--Boltzmann codes. We show how the effect of radiation can instead be consistently calculated in Newtonian simulations using the actual initial matter density and velocity fields for CDM and baryons in GR. 

Our GR results still only hold to first order in the metric potentials. Further work is required to assess the impact of non-linear corrections in GR. Our approach, constructing Newtonian motion gauges in which to interpret the unmodified Newtonian simulations, provides such a framework in which to incorporate higher-order corrections. We now have second-order Einstein--Boltzmann codes \cite{Pettinari:2013he,Huang:2012ub,Su:2012gt} 
which can calculate the full GR metric potentials up to second order. In particular these codes are able to calculate the GR matter bispectrum at leading order \cite{Tram:2016cpy}. In principle this can be used to set consistent relativistic initial conditions for Newtonian simulations up to second order, but only if we can develop a consistent GR interpretation of Newtonian simulations; this remains an outstanding challenge.

It is interesting to note that the definition of the Newtonian-motion gauges does not specify the underlying physics of the relativistic theory. While we focused on a Newtonian motion gauge to include the impact of radiation in GR, we could also construct a Newtonian motion gauge that includes more additional physics in the relativistic equations of motion, such as, for example, near-massless neutrinos, long-range dark matter interactions, or modified gravity scenarios.   
The modifications can be included in the Boltzmann code rather than in the Newtonian simulations so long as 
the new physics can be described by modifications to the linear evolution on large  scales.
Similarly, it may be possible to relate Newtonian motion gauges for different cosmological parameters or physical theories to the same Newtonian simulation. In this way, the numerical cost for parameter scans could be significantly reduced.

\acknowledgments

We thank Marco Bruni, Samuel Flender, Oliver Hahn, Christophe Ringeval and Dominik Schwarz for useful discussions.
CF is supported by the Wallonia-Brussels Federation grant ARC11/15-040 and the Belgian Federal Office for Science, Technical \& Cultural Affairs through the Interuniversity Attraction Pole P7/37.
CR acknowledges the support of the individual fellowships RA 2523/1-1 and RA 2523/1-2 from the Deutsche Forschungsgemeinschaft (DFG).
TT, RC, KK, and DW are supported by the UK Science and Technologies Facilities Council grants ST/K00090X/1 and ST/N000668/1.  
KK is supported by the European Research Council through grant 646702 (CosTesGrav).

\appendix
\section{The spatial gauge condition}\label{app:A}
As discussed in section \ref{NMgauge}, there exists a residual spatial gauge freedom which allows us to define a Newtonian motion gauge for any initial density and initial velocity.   This freedom is comparable to the residual gauge freedom in the synchronous gauge (see for example, ref.\,\cite{Malik:2008im}.)

A priori we have the freedom to impose any initial condition (or equivalently, any initial spatial gauge) in the Newtonian simulation. However, given that we are using a linear Einstein--Boltzmann solver, 
%
the choice of initial spatial gauge must be performed with caution. For example, if we were to choose Lagrangian (comoving) coordinates to absorb the full motion of particles into the space-time corresponding to a homogenous initial density (zero displacements) in Newtonian  
simulations, the metric would become non-linear at late times.
Eventually (linear) perturbation theory and thus our framework will break down.
 
To find a suitable choice for the initial gauge fixing, we consider the metric potential $H_{\rm T}$, which is fully specified by a choice of the spatial gauge transformation (\ref{eq:generalgaugetrafoHT}). In a radiation-free Universe (with only pressureless matter and $\Lambda$) the difference between the Newtonian potential and the Bardeen potential is due to the volume deformation and the relativistic correction of the Poisson equation~(\ref{eq:EE2}):
\be
k^2(\Phi^{\rm N} -\Phi) = 4\pi G a^2 \bar \rho_{\rm cdm} \left[ 3 H_{\rm L} -  3 {\cal H} k^{-1} \left( v - B \right)\right] = 4\pi G a^2 \bar \rho_{\rm cdm} \left(3\zeta - H_{\rm T}\right) \,,
\ee 
where in the second equality we have used the comoving curvature perturbation defined in eq.~(\ref{def:curvature}).
The Newtonian motion condition~(\ref{eq:NMgauge}) then reads
\be
\left( \partial_\tau + \mathcal{H} \right) \dot{H}_{\rm T} =-4\pi G a^2 \bar \rho_{\rm cdm} \left(3\zeta - H_{\rm T}\right) \,.
\ee
This equation shows that for given initial conditions for $H_{\rm T}$ and $\dot{H}_{\rm T}$ (equivalent to fixing $L$ and $\dot{L}$), 
$H_{\rm T}$ either converges towards $3\zeta$, where the right hand side vanishes, or it grows indefinitely.

As $\zeta$ is constant for a radiation-free cosmology, one particularly simple choice of initial spatial gauge conditions is $H_{\rm T}(\tau_{\rm ini}) = 3\zeta$ and $\dot{H}_{\rm T}(\tau_{\rm ini}) = 0$, resulting in a constant $H_{\rm T} = 3\zeta$ in this case, thereby avoiding the growth of $H_{\rm T}$.
This particular choice for the initial spatial gauge is identical to the $N$-body gauge spatial gauge condition. 
In the absence of radiation the spatial gauge then remains identical to the spatial gauge in the previously defined $N$-body gauge. 

In the presence of radiation, the dynamics is more complex and the metric solution is no longer so simple. However, we expect that radiation only adds small corrections to the metric potentials and when starting from the $N$-body gauge initial conditions, we expect the metric potentials of the Newtonian motion gauge remain close to the $N$-body gauge metric potentials at all times (as we have shown explicitly in section \ref{sec:post-processing}).  

Therefore, we conclude that the choice of $H_{\rm T}(\tau_{\rm ini}) = 3\zeta$ and $\dot{H}_{\rm T}(\tau_{\rm ini}) = 3\dot{\zeta}$, equivalent to the spatial gauge of the $N$-body gauge, is a good spatial gauge fixing for any Newtonian motion gauge, independent of the temporal gauge condition chosen.

\section{The temporal gauge condition}\label{app:Poisson}

For most of this paper we made use of a comoving orthogonal temporal gauge ($v=B$). In this appendix we will investigate a different temporal gauge fixing, providing a Newtonian motion gauge that offers a simple description (and generalisation) of the approach discussed by Chisari and Zaldarriaga \cite{Chisari:2011iq}.

If we identify the temporal coordinate with the well known Poisson gauge time, then the temporal gauge condition is\footnote{Note that in the typical gauge fixing of the Poisson gauge, $B=0$, is not a pure temporal gauge fixing. However, when setting both $H_{\rm T} = 0$ and $k B = \dot{H}_{\rm T}$, then $B=0$ follows as a consequence. Here we do not set $H_{\rm T} = 0$.} 
\be \label{eq:Poissongauge}
k B = \dot{H}_{\rm T}\,.
\ee 
The spatial gauge is fixed by the Newtonian motion condition~(\ref{eq:NMgauge})
and the initial spatial gauge choice, following the discussion in the previous appendix: $H_{\rm T}(\tau_{\rm ini}) = 3\zeta$ and $\dot{H}_{\rm T}(\tau_{\rm ini}) = 3\dot{\zeta}$. 
We therefore start the simulation in the $N$-body spatial gauge coordinates, while using the Poisson gauge time coordinate.

Using this temporal gauge we obtain the Einstein equations:
\begin{subequations}
 \begin{align}
 	k^2 \Phi &=  4 \pi G a^2 \left[ \bar \rho \delta + 3 {\cal H}  \left( \bar \rho+ \bar p \right) k^{-1} \left( v - B \right) \right] \,, \label{eq:ENP2} \\ 
 	k^2 \left( A + \Phi \right)  &= - 8 \pi G a^2 \bar p \Pi \,, 
	\label{eq:ENPA} 
 \end{align} 
while the fluid equations for the CDM component read:
\begin{align} 
\dot{\delta}_{\rm{cdm}} + k v_{\rm cdm} &= - 3 \dot H_{\rm L} \,,  \\  
\left[ \partial_\tau + {\cal H} \right] v_{\rm{cdm}} &= -k (\Phi + \gamma) 
 \,.
\end{align}
\end{subequations}
These equations result in Newtonian dark matter motion 
when expressed in terms of $\delta^{\rm N}_{\rm cdm}$ defined in~(\ref{def:Ndensity}) and $\Phi^{\rm N}$ defined in~(\ref{def:NPhi});
again, these dynamics are completely independent of the temporal gauge. However, since we use a different time coordinate, the underlying metric potentials are different in this temporal gauge.  
 
Instead of repeating the analysis using the new temporal gauge condition, we utilize the results computed in section \ref{sec:nbodyNM} to construct the space-time of this Newtonian motion gauge.
\begin{itemize}

\item The metric potential $H_{\rm T}$ is independent of the temporal gauge choice. The values for $H_{\rm T}$ computed in section \ref{sec:post-processing} are thus valid for any Newtonian motion gauge using initial conditions in the $N$-body gauge. 

\item The lapse function $A$ only depends on the temporal gauge choice. Since we employ the temporal gauge fixing of the Poisson gauge, the lapse can directly be identified with the well-known Poisson gauge lapse function $\Psi$ defined in eq.\,(\ref{def:Psi}). 

\item The shift $B$ is set directly by the temporal gauge condition, equation (\ref{eq:Poissongauge}).  
\item $H_{\rm L}$ can be constructed using the Einstein equations. The Bardeen potential is computed as:
\be
k^2 \Phi = 4 \pi G a^2 \left[ \bar \rho \delta + 3 {\cal H}  \left( \bar \rho+ \bar p \right) k^{-1} \left( v - B \right) \right] \,.
\ee
Given $\Phi$ and $H_{\rm T}$ we directly obtain $H_{\rm L}$ from (\ref{def:Bardeen}): 
\be
H_{\rm L} = \Phi - \frac 1 3 H_{\rm T} \,.
\ee
\end{itemize}

Using $H_{\rm T}$,
the lapse function $\Psi$ and the Bardeen potential $\Phi$, we can directly construct the space-time of the Newtonian motion gauge using the Poisson gauge time coordinate. 
In this space-time we can obtain a consistent relativistic interpretation of a Newtonian simulation. Since we use the same temporal coordinate of the Poisson gauge, we can also compute a simple spatial transformation connecting this gauge to the well known Poisson gauge using eq.\,(\ref{eq:generalgaugetrafoHT}):
\be
kL= -H_{\rm T} \,.
\ee
Thus, the output of Newtonian simulations can be understood in terms of the Poisson gauge \cite{Chisari:2011iq}, provided that 
the final positions of the CDM particles are displaced in the simulation according to
\be \label{eq:poisson}
  \fett{x}_{\rm Poisson} = \fett{x}_{\rm sim} + \nabla H_{\rm T} \,.
\ee

This is to be compared to the approach of Chisari and Zaldarriaga \cite{Chisari:2011iq}, who focused on a Poisson gauge interpretation of Newtonian simulations in a radiation-free universe.  
They required particles to be initially placed away from their positions in the Poisson gauge and instead effectively used the $N$-body gauge initial conditions.
During the $N$-body simulation they see Poissonian gauge terms cancelling in the evolution.  At the end, they give the particles a final displacement of ${\nabla}(3\zeta)$.   
 
The Newtonian motion gauge provides a clear interpretation of this approach.  Using $N$-body gauge initial conditions turns out to be crucial, as Poisson gauge initial conditions would result in a rapidly growing value for $H_{\rm T}$ (see previous appendix).   Their final displacements can be seen to shift from the Newtonian motion gauge to the Poisson gauge. 
In the limit of vanishing anisotropic stress and pressure perturbations, our result simplifies and we see $H_{\rm T} = 3\zeta$; thus their displacement agrees with the general gauge transformation to the Poisson gauge described in eq.\,(\ref{eq:poisson}).   However, our more general method can be applied even when radiation corrections cannot be ignored.

\section{Total matter gauge and \texorpdfstring{$N$}{N}-body gauge in the limit of vanishing radiation}\label{app:tom}

In a Universe composed of only dark matter and a cosmological constant we have shown that the $N$-body gauge has, to linear order, Newtonian equations of motion. However it is not the only gauge with this property; see for example the total matter gauge \cite{Malik:2008im,Flender:2012nq,Villa:2015ppa} defined by $v=B$ and $H_{\rm T} = 0$.\footnote{In ref.\,\cite{Flender:2012nq}, the considered gauge has been denoted with Newtonian matter gauge.} In this gauge, the relativistic Euler and Poisson equation take the Newtonian form, while we have a non-vanishing value of $H_{\rm L} \equiv \zeta$. Since the comoving curvature is conserved during matter/$\Lambda$ domination, this constant volume deformation does not affect the dark matter continuity equation. 
Thus, similar to the $N$-body gauge, we have Newtonian evolution equations for the dark matter particles. 

However, in addition to having Newtonian equations of motion,
the $N$-body gauge also has a vanishing volume deformation, $H_{\rm L} = 0$, allowing the direct identification of the simulation density with the relativistic density. On the other hand, in the total matter gauge the volume deformation makes it impossible to find consistent initial conditions for the Newtonian simulation \cite{Fidler:2015npa}: If particles are placed at their relativistic positions initially, the Newtonian simulation does not compute the relativistic density when embedding the particles in Newtonian flat space. As a consequence the simulation potential is not equal to the Bardeen potential and particles are not moved according to GR. 

The Newtonian dark matter dynamics in both the $N$-body gauge and the total matter gauge rely on a constant value of $\zeta$, which is the case in matter/$\Lambda$ domination. Beyond this limit, i.e., when radiation is still significant, we can employ the more general Newtonian motion gauges as presented in this paper.

\bibliographystyle{JHEP}
\bibliography{references}

\end{document}